\documentclass[12pt]{report}
\usepackage[dvips]{graphicx}
\newcommand{\sptwo}{1.4}

\newcommand{\doublespace}{\edef\baselinestretch{\sptwo}\Large\normalsize}

\begin{document}
\doublespace
\begin{flushright}
Contributed Talk at the 12th International Conference\\
Recent Progress in Many-Body Theories\\
August 23-27, 2004 Santa Fe, New Mexico
\end{flushright}
\begin{center}
{\bf  Strongly Interacting Fermi Gases of Atoms  Confined in a
Harmonic Trap}\\
  Yeong E. Kim and  Alexander L. Zubarev
 \\
Purdue Nuclear and Many-Body Theory Group\\
Department of Physics, Purdue University \\
West Lafayette, IN  47907-1397 \\
\end{center}

\begin{quote}
Dynamics of strongly interacting Fermi gases, consisting of a 50-50 mixture of
 two different fermionic species, is investigated. For the equation of state
 we consider a Pad\'{e} [2/2] approximations, which gives the weak-coupling perturbative formula (up to 4th order) in the low density regime, the unitary-limit Monte Carlo result in the high density regime, and reproduces the 4-fermion prediction for dimer-dimer scattering length in the BEC region. We use a time-dependent LDA to derive various properties of the Fermi gas under a harmonic confinement and compare them with the data of  very recent experiments of $^6Li$ atoms
 across a Feshbach resonance.
\end{quote}

\vspace{5mm}

\pagebreak

\noindent
In this talk the dynamics of strongly interacting  dilute Fermi gases
 (dilute in the
sense that
 the range of interatomic potential is small
compared with inter-particle spacing) consisting  of a 50-50 mixture of two
 different
 states and
confined in a harmonic trap
$V_{ext}(\vec{r})=(m/2)(\omega_{\perp}^2 (x^2+y^
2)+
\omega_z^2 z^2)$
 is investigated in the single equation approach to the time-dependent
density-functional theory [1,2]
$$
i\hbar \frac{\partial \Psi}{\partial t}=-\frac{ \hbar^2}{2 m} \nabla^2 \Psi
+V_{ext} \Psi+V_{xc}\Psi,
\eqno{(1)}
$$
where $V_{xc}(\vec{r},t)=[\frac{\partial n \epsilon(
n)}{\partial n}]_{n=n(\vec{r},t)}$, 
 the density of the system is
$
n(\vec{r},t)=\mid \Psi(\vec{r},t)\mid^2,
$
and the velocity field 
$
\vec{\mbox{v}}(\vec{r},t)=\hbar(\Psi^\ast \nabla \Psi-
\Psi \nabla \Psi^\ast)/(2 i m n(\vec{r},t)).
$

 Let us come back to the variational 
formulation of the Kohn-Sham time-dependent theory
$$
\delta \int dt<\psi|i\hbar \partial_t-H|\psi>,
\eqno{(2)}
$$
where $|\psi>$ is a product of two Slater determinants, one for each internal
 state built up by the Kohn-Sham orbitals $\psi_i$,  and $H=T+U$ is the LDA
Hamiltonian.

Eq.(1) can be derived from Eq.(2) using two approximations

(i) local transform $\psi_i\approx \phi_i \exp (i\hbar \chi/m)$,
where $\psi_i$ and $\chi$ are real functions,

\noindent
and

(ii) $<\phi|T|\phi>\approx \int (t_{TF}(n)+t_W(n))n(\vec{r},t)d^3 r,$

where $|\phi>$ is the product of two Slater determinants built on $\phi_i$ 
alone,
$t_{TF}(n)=(3 \pi^2)^{2/3} (3 \hbar^2/(10 m)) n(\vec{r},t)^{2/3}$ is the 
Thomas-Fermi kinetic energy density, and
$t_W(n)=(\hbar^2/(8m))(\nabla n)^2/n$ is the original von Weizs\"{a}cker
density (OWD).

We note here that the approximations (i) and
 $<\phi|T|\phi>\approx \int t_{TF}(n)n(\vec{r},t)d^3 r$ lead to the
 hydrodynamic approximation  (HA) [3].
But near the surface the Hartree-Fock (HF) type densities are
 proportional to the square of the last occupied state. Therefore, the OWD is
 important in this case and it is expected to determine the asymptotic
 behavior of the density at large distances. It is also expected that the OWD is
 important in the case of the tight radial trapping, $\lambda \ll 1$.

Taking into account that in the limit $a\rightarrow -0$, where $a$ 
is the scattering
 length,
 Eq.(2) leads to exact equations describing scaling properties of a ideal gas
 trapped in a time-dependent harmonic potential [4],
we  do expect that the Kohn-Sham approach may be an appropriate to describe
dynamics  of cold fermionic gas in the regime when superfluid quantum 
 hydrodynamics, Eq.(1) is
not
applicable.

For the negative S-wave scattering length between the two fermionic species,
 $a<0$,
in the low-density regime, $k_F\mid a \mid \ll 1$, the ground state energy per
particle , $\epsilon(n)$, is well represented by an expansion in power of 
$k_F \mid a \mid$ [5]
$$
\epsilon(n)=2 E_F[\frac{3}{10}-
\frac{1}{3 \pi} k_F \mid a \mid+0.055661 (k_F\mid a \mid)^2
-0.00914 (k_F\mid a \mid)^3+...],
\eqno{(3)}
$$
where $E_F=\hbar^2 k_F^2/(2 m)$.

In the opposite regime, $a\rightarrow - \infty$
,the Bertsch many-body problem, 
 $\epsilon(n)$ is proportional to that of the non-interacting Fermi
 gas
$$
\epsilon(n)=(1+\beta)\frac{3}{10} \frac{\hbar^2 k_F^2}{m},
\eqno{(4)}
$$
where a  universal parameter  $\beta$ [6]  is estimated to be $\beta=-0.56$ [7].

 In the $a\rightarrow +0$ limit the system reduces to the
dilute Bose gas of dimers [8]

$$
\epsilon(n)=E_F(-1/(k_F a)^2+a_m k_F/(6 \pi)+...),
\eqno{(5)}
$$
where $a_m$ is the boson-boson scattering length.
 Solution of 4-fermion problem for contact scattering provided the value
$a_m\approx 0.6 a$ [9].

In Refs.[1,2] it has been proposed a simple interpolation of the form
$\epsilon(n)\approx
E_F P(
k_F  a)$ with a smooth function $P(x)$ mediating between the known limits.

For the negative $a$ it has  been proposed a [2/2] Pad\'{e} approximant for the function
$P(x)$
$$
P(x)=\frac{3}{5}-2\frac{\delta_1\mid x \mid+\delta_2 x^2
}{1+\delta_3\mid x\mid+\delta_4 x^2},
\eqno{(6)}
$$
where $\delta_1=0.106103$, $\delta_2=0.187515$, $\delta_3=2.29188$,
$\delta_4=1.11616$.
Eq.(6) is constructed to reproduce the first four terms of the expansion (3) in
 the low-density regime and
 also to  reproduce
exactly results of
 the recent Monte Carlo calculations [7], $\beta=-0.56$, in the  unitary limit,
$k_F a \rightarrow -\infty$. 

 For the positive $a$ case ( the interaction is strong enough to form bound 
molecules with energy $E_{mol}$)  it has been considered a [2/2] Pad\'{e} 
approximant
$$
P(x)=\frac{E_{mol}}{2 E_F}+\frac{\alpha_1 x+\alpha_2 x^2}{1+\alpha_3 x+\alpha_4
x^2},
\eqno{(7)}
$$
where
 parameters $\alpha$ are fixed by two continuity conditions at large $x$,
$1/x\rightarrow 0$, and by two continuity conditions at small $x$.
For example, $\alpha_1=0.0316621$, $\alpha_2=0.0111816$, $\alpha_3=0.200149$,
and $\alpha_4=0.0423545$ for $a_m=0.6 a$.

The aspect ratio, presented in Fig. 1, shows that the effect of inclusion of the OWD (quantum pressure) on the expansion of superfluid for the conditions of Ref.[6] is about 1\%.

Fig. 2 and Fig. 3 show the comparison between  [2/2] Pad\'{e} approximations,
Eqs.(6,7),
and the  lowest order constrained variational (LOCV) approximation [10] and
 the BCS mean-field theory [11] for
$\epsilon(n)$.
The LOCV calculations  agree very well with the [2/2] Pad\'{e}
 approximation results on the BCS side ($a<0$). It is evident the difference
 between  our results and the BCS mean-field theory calculations. For example,
the BCS mean-field gives $\beta=-0.41$. 

The predictions of Eq.(1) with $\epsilon(n)$ from Eq.(6) for the axial cloud
 size of   strongly interacting $^6Li$ atoms are shown in Fig 4. It
indicates
that the TF approximation of the kinetic energy density is a very good
approximation for the experimental conditions of Ref.[12], $N \lambda \approx
10^4$  (inclusion of
 the OWD gives a negligible effect, $<0.5\%$).

In Fig. 5, we present the calculations for the frequency of the radial
compression mode $\omega_{rad}$ as a
function
 of the dimensional parameter $(N^{1/6} a/a_{ho})^{-1}$ in the case of an 
anisotropic trap ($\omega_x=\omega_y=\omega_{\perp}$,
$\omega_z/\omega_{\perp}=\lambda$). One can easily see that the corrections to
the
  hydrodynamic
approximation
(HA) (inclusion of the OWD)
 are important
 even
 for relatively large $N$ and $\lambda N$. For example,
the correction to  $\omega_{rad}$ in unitary limit is larger than 11\% and 25\%
for $\lambda=10^{-2}$, $N=10^4$ and $\lambda=10^{-2}$, $N=10^3$,
 respectively.

In the HA, $\omega_{rad}$ is independent of $N$ for a fixed
$(N^{1/6} a/a_{ho})^{-1}$. The deviation from this behavior does not
demonstrate the cross-over to the $1D$ behavior, since $\lambda N>1$.
It demonstrates that the validity of the HA depends on the properties of the
 trap. We note here that the collective modes of the Fermi gas under harmonic 
confinement in the framework
 of the hydrodynamic approximation was  considered recently in [13,14].

In Fig.6, the calculated radial compressional frequency is compared with 
experimental data [15] in the BCS-BEC crossover region. There is a very good agreement
 between  calculations and experimental data [15].

 However our calculations for $
\omega_{rad}$ disagree with experimental data of Ref.[16].
It is well known that the hydrodynamic equation is expected to be applicable
for describing the macroscopic excitations of the system up to energies of the
 order of the energy gap, $\Delta$, needed to break-up a Cooper pair. But for
 the trapped gas $\Delta$ is a function of $\vec{r}$ ($\Delta$ decreasing when
 we
 go away from the center). It is naturally to assume that
 condition of the
applicability of hydrodynamics to describe the macroscopic excitations of the
system at $T=0$ is
$$
\frac{\hbar \Omega}{\tilde{\Delta}}\ll 1,
\eqno{(8)}
$$
 where $\Omega$ is the frequency of the macroscopic excitations,
$\tilde{\Delta}=\int n(\vec{r}) \Delta(\vec{r}) d\vec{r}/N$.
To calculate $\tilde{\Delta}$ we have used results of Refs.[17,18].

Fig.7 shows the comparison between the average energy gap and frequencies of the transverse and axial breathing modes.
It may explain why our calculations disagree with experimental data of
 Ref.[16], see also Ref.[19]. Taking into account the trap difference between
 Ref.[15] and
 Ref.[16] we expect that the Duke University group will reproduce the 910G
 strong change in the $\Omega_{rad}$ of
 Ref.[16] at $B\approx 1000G$.

As for axial mode,  interesting results may start at $B\approx 1250$G if
$\omega_{ho}$ of the trap is about  $2 \pi 60$ Hz.\\

We thank J.E. Thomas for his interest and providing us with the experimental
 data.
\pagebreak
\begin{figure}[ht]
\includegraphics{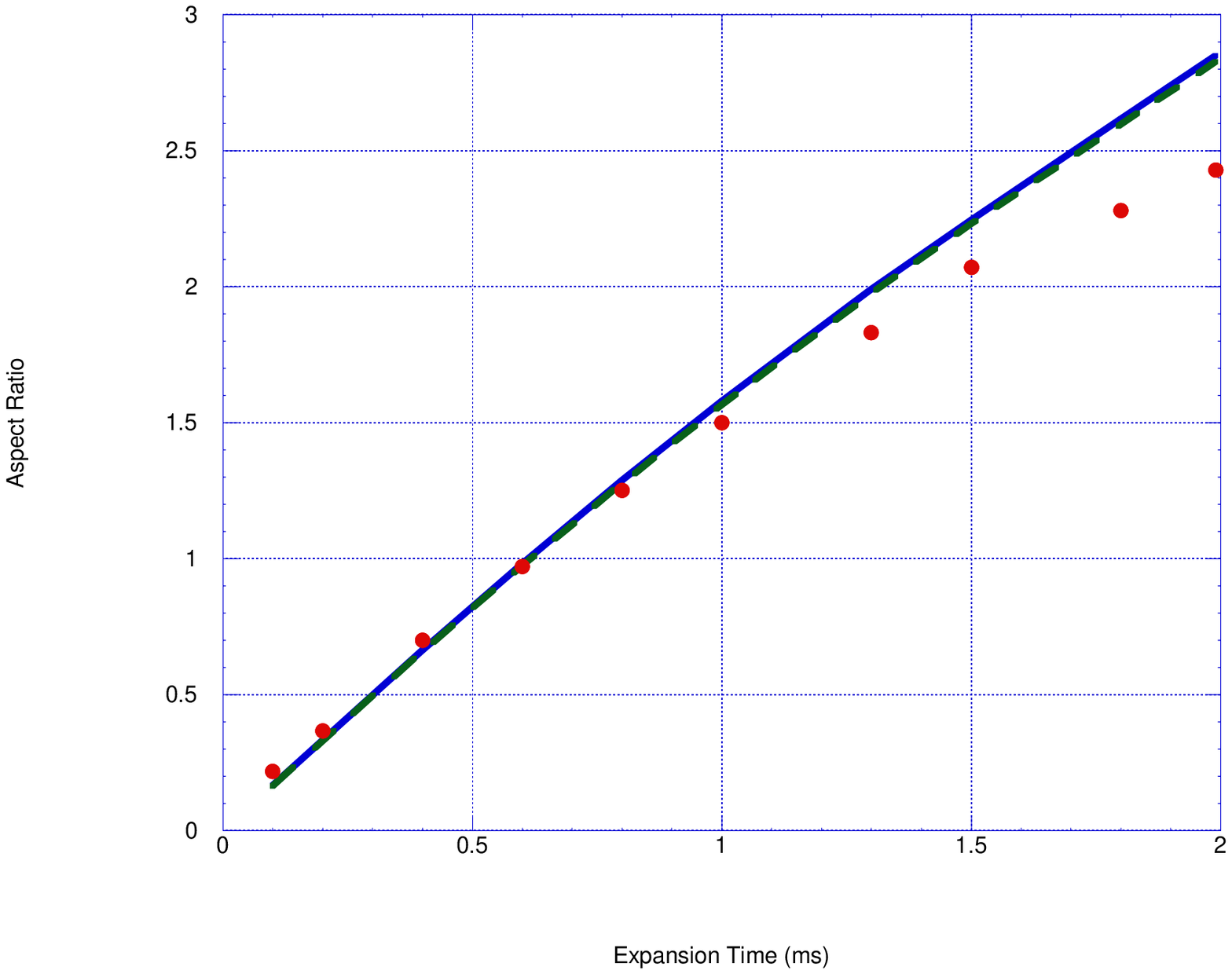}
\end{figure}
Fig. 1. Aspect ratio of the cloud of the $N=7.5 \times 10^4$ $^6Li$ atoms as a
function of time
 after release from the trap ($\omega_{\perp}=2 \pi \times 6605$Hz,
$\omega_z=2 \pi \times 230$Hz).
The circular dots indicate experimental data from the Duke University group [6].
The solid line and the dashed line represent theoretical calculations in the
 unitary limit ($a\rightarrow -\infty$) including the quantum pressure term
  and in  the  hydrodynamic approximation,
 respectively.

\pagebreak
\begin{figure}[ht]
\includegraphics{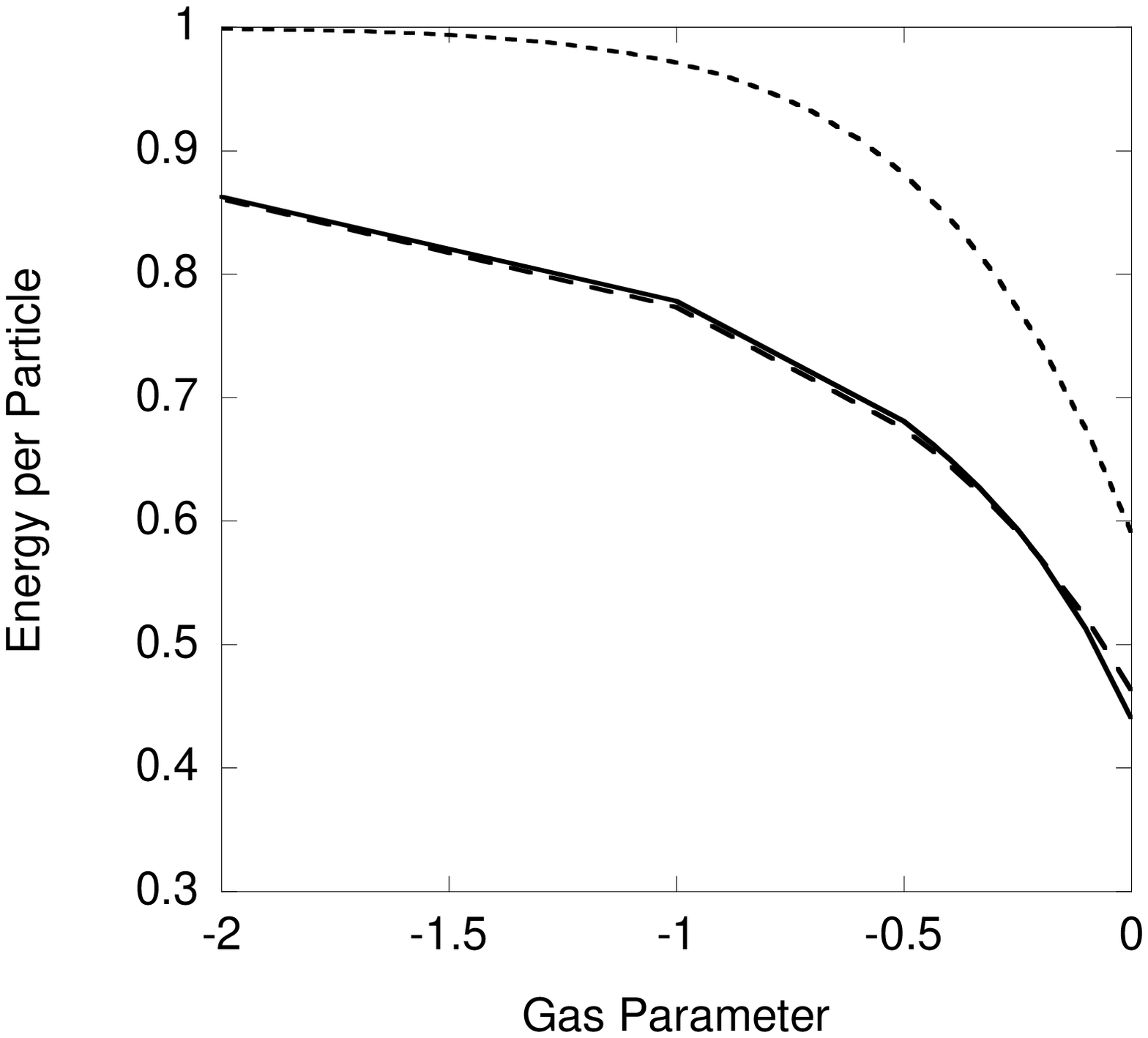}
\end{figure}
Fig.2. The ground state energy per particle, $\epsilon(n)$,
  in units of $3 \hbar^2 k_F^2/(10 m)$ as a function of the gas  parameter
 $(k_F a)^{-1}$. The solid line,
the long dashed line  and the short dashed line
 represent the results calculated
using the [2/2] Pad\'{e} approximation, Eq.(6), the LOCV approximation,
 and the BCS mean-field theory,  respectively.

\pagebreak
\begin{figure}[ht]
\includegraphics{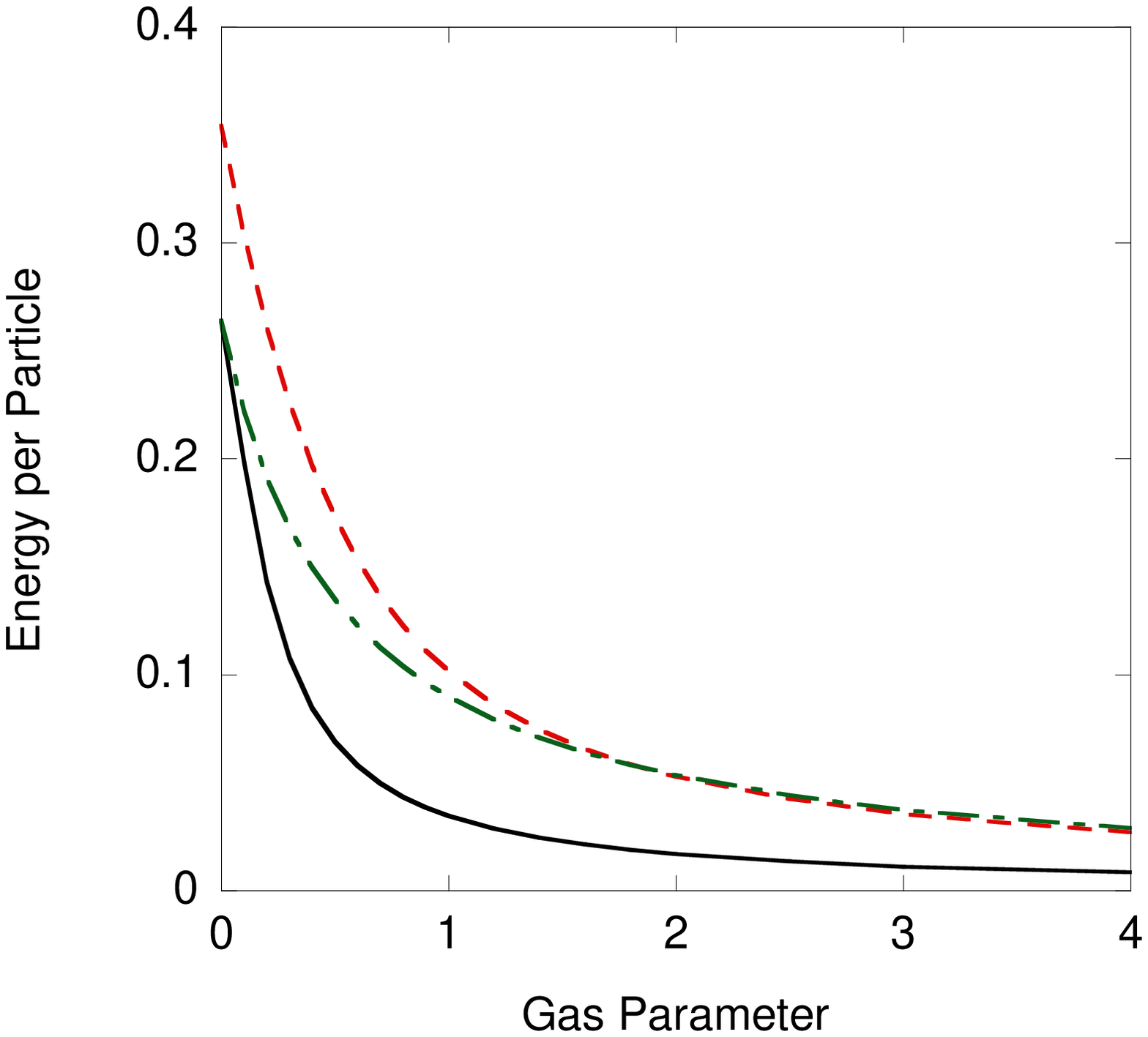}
\end{figure}
Fig.3. The ground state energy per particle, $\epsilon(n)+|E_{mol}|/2$,
  in units of $\hbar^2 k_F^2/(2 m)$ as a function of the gas parameter
 $(k_F a)^{-1}$.
The dashed line,  the dotted-dashed line and
the solid line
represent the results calculated using the BCS mean-field theory, the [2/2]
Pad\'{e}
 approximation, Eq.(7), with $a_m=2a$, and $a_m=0.6a,$ respectively.

\pagebreak
\begin{figure}[ht]
\includegraphics{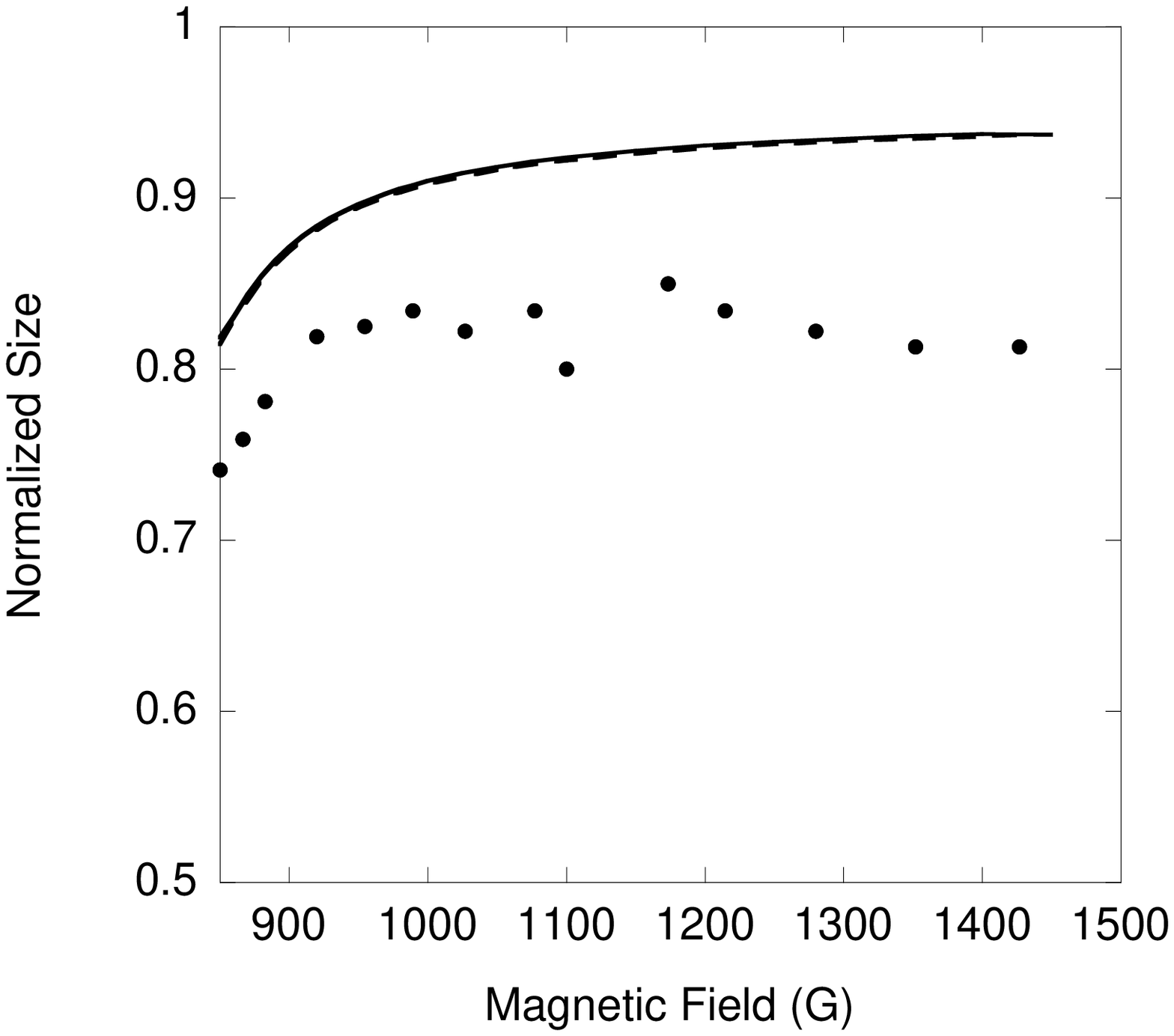}
\end{figure}
Fig. 4. Axial cloud size of strongly interacting $^6Li$ atoms
 after normalization to a non-interacting Fermi gas
with $N=4 \times 10^5$ atoms as a function of the magnetic field $B$.
 The trap parameters are $\omega_{\perp}=
2 \pi \times 640$Hz, $\omega_z=2 \pi (600 B/kG+32)^{1/2}$Hz.
 The solid line and dashed line  represent the results of
 theoretical calculation that includes the OWD or uses the TF approximation
 for the kinetic energy density, respectively. The circular dots indicate
 experimental data from the Innsbruck group [12].

\pagebreak
\begin{figure}[ht]
\includegraphics{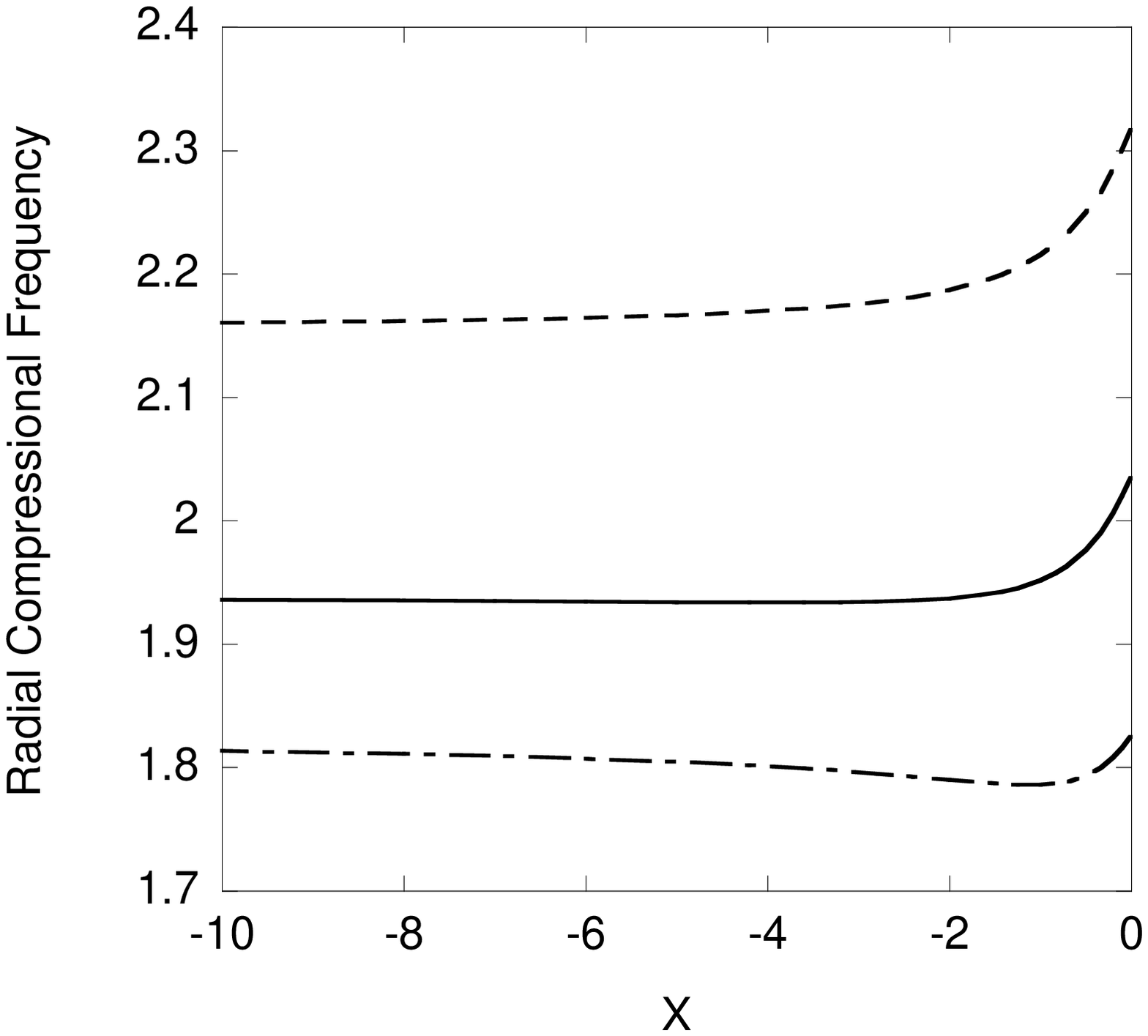}
\end{figure}
Fig. 5.  Radial compressional frequency, $\omega_{rad}$, of the cloud of the $N=
10^4$ fermions
(solid line) and $N=10^3$ fermions (dashed line)
 in unit of $\omega_{\perp}$ as a function
 of the dimensional parameter X=$(N^{1/6} a/a_{ho})^{-1}$. The trap parameter
$\lambda$ is assumed to be equal to $10^{-2}$. The lower line (dashed-dotted
 line)
 represents the results
in the   hydrodynamic
approximation,  in which  $\omega_{rad}$ is independent of $N$
for a fixed $(N^{1/6} a/a_{ho})^{-1}$.

\pagebreak
\begin{figure}[ht]
\includegraphics{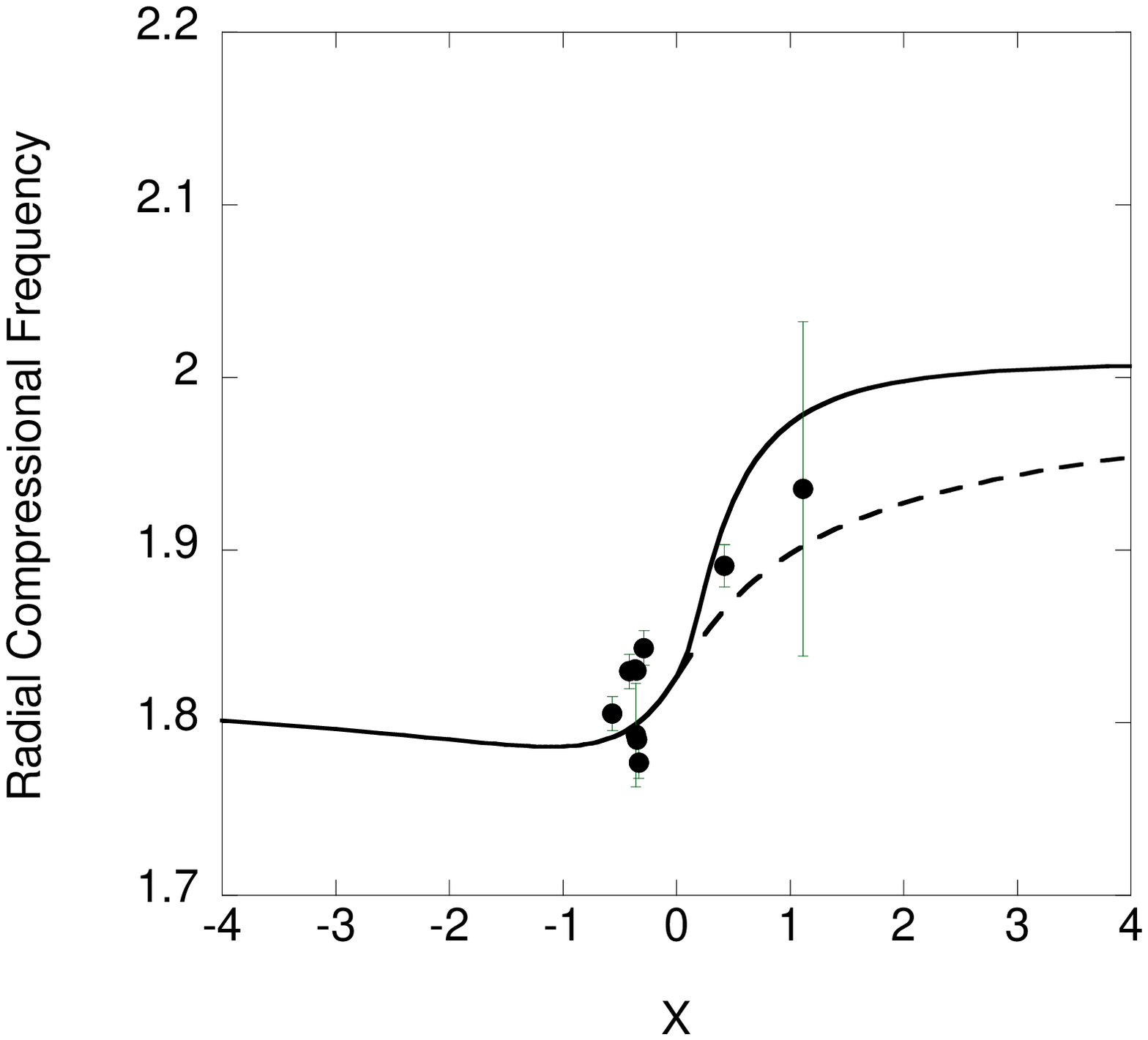}
\end{figure}
Fig.6. The radial compressional frequency as  a function of X=$(N^{1/6}a/a_{ho})
^{
-1}$. The solid line and the dashed line represent the results
calculated using the [2/2] Pad\'{e}
 approximation
with $a_m=0.6 a$
and $a_m=2 a$, respectively. The solid circles with error bars are the
experimental results given by the Duke University group [15].

\pagebreak
\begin{figure}[ht]
\includegraphics{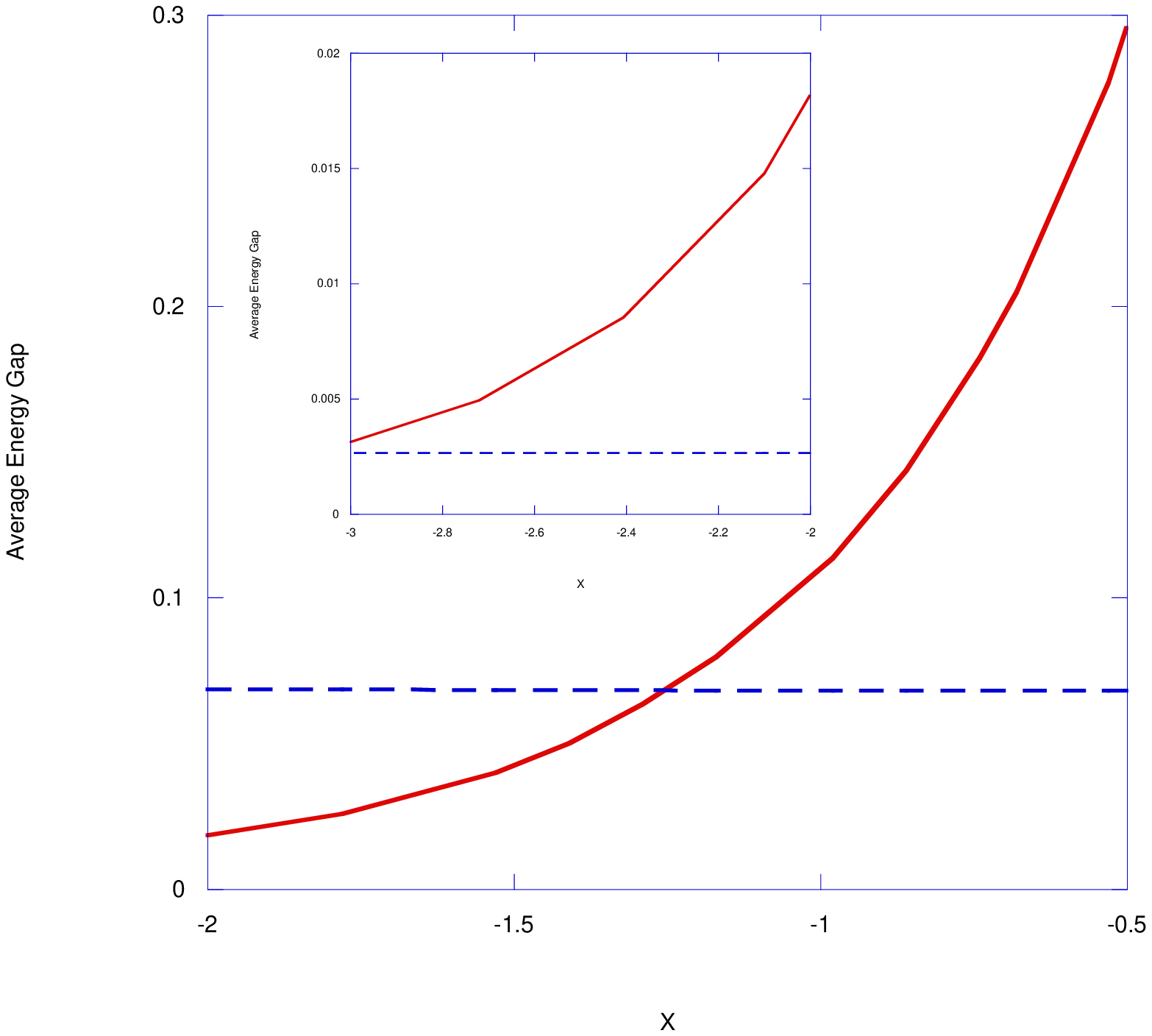}
\end{figure}
Fig.7. The average energy gap in units of $\hbar \omega_{ho} N^{1/3}$ of  an 
elongated
trapped $N=4 \times 10^5$  Fermi atoms ($\lambda=0.045$) as a function of the
parameter $X=(k_F(0) a)^{-1}$ (solid lines). The dashed lines in the main plot
 and in the inset are the frequencies of the transverse and axial breathing
 modes in the same units ($\hbar \omega_{ho} N^{1/3}$), respectively.

\pagebreak
{\bf References}

\noindent
[1] Y.E. Kim and A.L. Zubarev, Phys. Lett. A327, 397 (2004).

\noindent
[2] Y.E. Kim and A.L. Zubarev, Phys. Rev. A (in print); cond/mat/0404513.

\noindent
[3] L. Pitaevskii and S. Stringari, {\it Bose-Einstein Condensation},
(Clarendon Press, Oxford, 2003).

\noindent
[4] G.M. Bruun and C.W. Clark, Phys. Rev. A{\bf 61}, 061601 (2000).

\noindent
[5]  G.A. Baker, Jr., Int. J. Mod. Phys. B{\bf15}, 1314 (2001); Phys. Rev. C{\bf
60},
054311 (1999) and references therein.

\noindent
[6] K.M. O'Hara, S.L. Hemmer, M.E. Gehm, S.R. Granade, and  J.E. Thomas,
Science {298}, 2179 (2002).

\noindent
[7] J. Carlson, S.-Y. Chang, V.R. Pandharipande, and K.E. Schmidt,
Phys. Rev. Lett. {\bf
 91},
050401 (2003).

\noindent
[8] A.J. Leggett, in {\it Modern Trends in the Theory of Condensed
Matter},
 edited by A. Pekalski and R. Przystawa, Lecture Notes in Physics
Vol. 115 (Springer-Verlag, Berlin, 1980) pp. 13-27;
P. Nozi\`{e}res and S. Schmitt-Rink, J. Low. Temp. Phys. {\bf 59}, 195 (1985).

\noindent
[9] D.S. Petrov, C. Salomon and G.V. Shlyapnikov, cond-mat/0309010.

\noindent
[10]  V.R. Pandharipande, Nucl. Phys. A{\bf 174}, 641 (1971);
S. Cowel, H. Heiselberg, I.E. Morales, V.R. Pandharipande, and C.J. Pethick,
 Phys. Rev. Lett. {\bf 88}, 210403 (2002); H. Heiselberg, J. Phys.  B{\bf 37},
S141 (2004); H. Heiselberg, cond-mat/0403041.

\noindent
[11] Hui Hu et al, cond-mat/0404012

\noindent
[12] M. Bartenstein, A. Altmeyer, S. Riedl, S. Jochim, C. Chin, J.
Hecker Denschlag, and R. Grimm, Phys. Rev. Lett. {\bf92}, 120401 (2004).

\noindent
[13] A. Bulgac and G.F. Bertsch, cond-mat/0404687.

\noindent
[14] N. Manini and L. Salasnich, cond-mat/0407039 and references therein.

\noindent
[15] J. Kinast, S. L. Hemmer, M. E. Gehm, A. Turlapov, and J. E. Thomas,
Phys. Rev. Lett. {\bf 92}, 150402 (2004).

\noindent
[16]  M. Bartenstein, A. Altmeyer, S. Riedl, S. Jochim, C. Chin,
 J. Hecker Denschlag, and R. Grimm,  Phys. Rev. Lett. {\bf 92}, 203201 (2004).

\noindent
[17]  L.P. Gorkov and T.K. Melik-Bakhudarov, Sov. Phys. JETP, {\bf 13}, 1018 (1961
).

\noindent
[18] S.Y. Chang, V.R. Pandharipande, J. Carlson and K.E. Schmidt,
 physics/0404115; G.E. Astrakharchik et al, cond-mat/0406113.

\noindent
[19] R. Combescot and X. Leyronas, cond-mat/0405146.

\end{document}